# Magnetic proximity effect at the interface of two-dimensional materials and magnetic oxide insulators

Junxiong Hu[a,b], Jiangbo Luo[a,c], Yuntian Zheng[a], Jiayu Chen[a], Ganesh Ji Omar[a], Andrew Thye Shen Wee[a], A. Ariando[a,*]

[a]Department of Physics, National University of Singapore, 2 Science Drive 3, Singapore 117542, Singapore
[b]Centre for Advanced 2D Materials, National University of Singapore, Singapore 117546, Singapore
[c]College of Chemistry and Chemical Engineering, Hunan University, Changsha 410082, P. R. China
*E-mail address: ariando@nus.edu.sg

**ABSTRACT**

Two-dimensional (2D) materials provide a platform for developing novel spintronic devices and circuits for low-power electronics. In particular, inducing magnetism and injecting spins in graphene have promised the emerging field of graphene spintronics. This review focuses on the magnetic proximity effect at the interface of 2D materials and magnetic oxide insulators. We highlight the unique spin-related phenomena arising from magnetic exchange interaction and spin-orbital coupling in 2D materials coupled with magnetic oxides. We also describe the fabrication of multifunctional hybrid devices based on spin transport. We conclude with a perspective of the field and highlight challenges for the design and fabrication of 2D spintronic devices and their potential applications in information storage and logic devices.

**Contents**

## 1. Introduction

"The interface is the device." This famous phrase was coined by Nobel laureate Herbert Kroemer for the development of semiconductor heterostructures used in high-speed electronics and optoelectronics, which started more than 50 years ago [1]. Now we are again at the center of a similar revolution, but this time at the interfaces of oxide heterostructures and two-dimensional (2D) van der Waals heterostructures [2, 3]. Oxide heterostructures have mesmerized the scientific community in the last decade due to the possibility of creating tunable multifunctionalities such as metallicity, ferroelectricity, ferromagnetism, and superconductivity at a single interface [4-6]. Furthermore, the discovery of 2D materials and van der Waals heterostructures allows the combination of different 2D materials into functional stacks, allowing the fabrication of a wide range of interesting electronic and optoelectronic devices, resulting in the observation of exciting quantum phenomena [7-10]. Recently, twist angle engineering between 2D crystal stacks (so-called twistronics) has added a new degree of freedom for tuning the electronic states of the heterostructures [11-13]. Inspired by the rich properties of functional oxides and 2D van der Waals heterostructures, combining 2D materials and functional oxides in hybrid heterostructures may lead to novel phenomena and intriguing physics, creating potential applications for future (multi-)functional devices [14, 15].

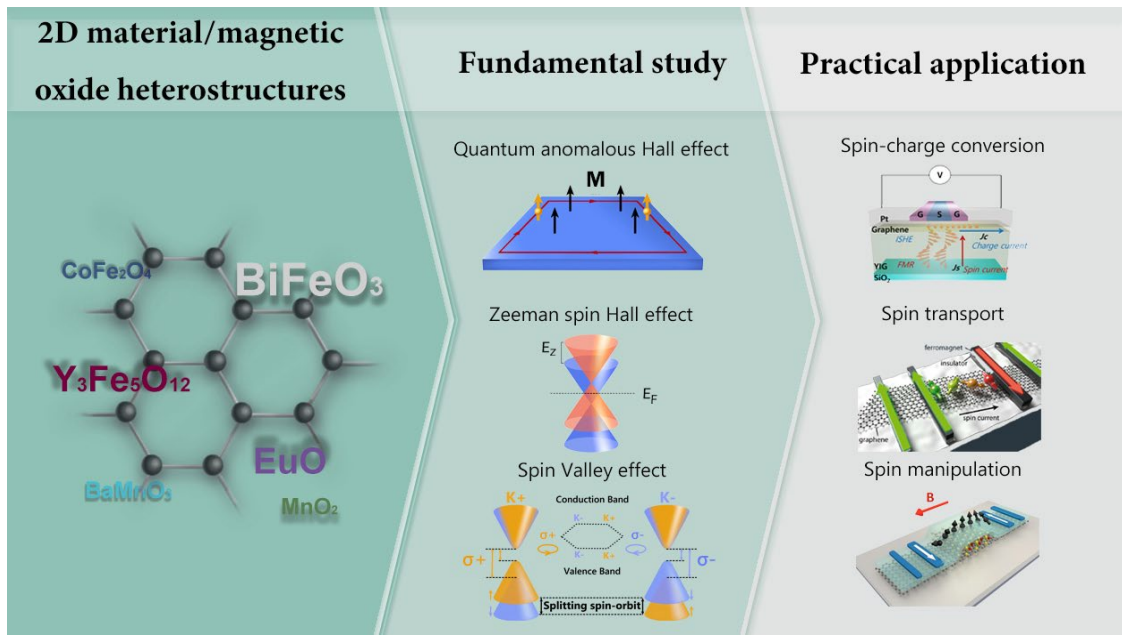

**Fig. 1.** Overview of the research progress at the interface of graphene and magnetic oxide insulators. The fundamental studies in graphene/magnetic oxide heterostructures primarily focus on the quantum anomalous Hall effect, Zeeman spin Hall effect, and spin valley effect. While for practical applications, the progress of graphene spintronics primarily relies on spin-charge conversion, spin transport, and spin manipulation. Figure adapted with permission from: ref [25-27], American Physical Society.

One mainstream class of hybrid heterostructure combines 2D materials and magnetic oxides for studying magnetic proximity effect (MPE) at their interface. The MPE, induced by an exchange-interaction at the interface of two neighboring layers, has been used to manipulate spintronic [16], superconducting [17], and topological phenomena [18] in heterostructures. This effect is susceptible to interfacial electronic properties such as electron wavefunction overlap and band alignment, leading to proximity-induced changes in the electrical, optical and spin-related properties [19]. Owing to the short range of the magnetic exchange and spin-orbit interactions, the atomically thin nature of 2D materials promotes the design of artificial intelligence materials and innovative devices by proximity effects [20, 21].

There are two primary motivations for inducing magnetism and injecting spins in graphene, as illustrated in Fig. 1. The first is to realize Chern insulating states, which exhibit the long-sought quantum anomalous Hall effect (QAHE) [22]. In 1988, Haldane theoretically proposed that a 2D honeycomb structure can be a topological quantum state that is insulating in bulk, but hosts robust conducting edge states and exhibits quantized Hall conductivity, even in the absence of an external magnetic field, so-called the QAHE [23]. Graphene is a unique two-dimensional electron system because of its ability to maintain electrical isolation and Dirac band dispersion and is an ideal prototype material for realizing the QAHE [24]. The second goal is to realize the generation, transport, and manipulation of spin signals in graphene [25-27]. Owing to its small spin–orbit coupling and hyperfine interaction, graphene is expected to possess a long spin relaxation time and long length even at room temperature [28, 29]. Moreover, graphene has a unique Dirac band structure, gate-tunable carrier density and high carrier mobility, making it attractive for 2D spintronics [30, 31].

However, graphene is not magnetic in its pristine form and has extremely weak intrinsic spin-orbit coupling [32, 33]. To realize the QAHE and spin transport in graphene, both magnetism and spin-orbit coupling (SOC) should be induced externally. Various efforts have been devoted to inducing magnetism and SOC in graphene, such as by atom doping [34], edge engineering [35], and also the MPE [36]. Compared with other methods, the advantage of MPE is that it is easy to control the interface and minimally degrade the quality of graphene, and thus much effort has been devoted to the development of the proximity effects at the interface of graphene and magnetic substrates. Examples include ferromagnets such as EuS and EuO [37, 38], ferrimagnets such as $Y_3Fe_5O_{12}$ (YIG) and $CoFe_2O_4$ (CFO) [39, 40], and antiferromagnets $BiFeO_3$ (BFO) and CrSe [36, 41].

Compared with other magnetic substrates, magnetic oxides have various advantages for MPE. First, benefitting from advanced thin-film growth techniques such as pulsed laser deposition (PLD) or molecular beam epitaxy (MBE), atomic-level control of oxide heterostructures has become possible [42]. The atomically flat interface is a prerequisite for MPE. Second, magnetic oxides have rich magnetic properties, which can be tuned by the substrate lattice, oxygen vacancies, and the growth dynamics such as pressure and temperature [43]. Third, most oxides are insulating and have large bandgaps, ensuring graphene is the only transport channel [38]. These promising features make magnetic oxides an ideal platform for MPE in graphene.

This review focuses on the MPE between 2D materials and magnetic oxides. First, we discuss the theoretical and experimental progress made on various 2D materials/magnetic oxide heterostructures, especially the band alignment, the spin splitting energy, and band structure. We also present various developed spintronics devices. Finally, we discuss potential applications and future research directions and perspectives.

## 2. MPE at the interface of graphene and magnetic oxide insulators

2D materials like graphene are expected to experience strong MPE in heterostructures with magnetic oxide insulators, owing to the short range of the magnetic exchange and spin-orbit interactions [37]. MPE can be significantly induced in non-magnetic 2D materials by a magnetic substrate, leading to the change of band structure and properties of 2D materials. The induced magnetism is characterized by a net local spin polarization and spin splitting bands. In the absence of SOC, the split bands are equal for the two valleys in graphene [20]. The strength of MPE can be parameterized by the exchange coupling strength $\lambda_{EX}$ (Table 1). The goal of MPE in graphene is to achieve a large $\lambda_{EX}$ while maintaining the spin transport capabilities of the isolated graphene.

Table 1: Summary of exchange coupling strength of graphene on magnetic oxides.

| 2D materials | Magnetic oxide | Magnetic property | Exchange coupling strength $\lambda_{EX}$ | Methods |
|---|---|---|---|---|
| graphene | **BiFeO$_3$** | Antiferromagnet insulator | 70 meV | Theory, DFT [36] |
| graphene | **BiFeO$_3$** | Antiferromagnet insulator | 2.29-33.37 meV* | Exp, transport [51] |
| graphene | **Y$_3$Fe$_5$O$_{12}$** | Ferrimagnet insulator | 52 meV (electron) 115 meV (Holes) | Theory, DFT [39] |
| graphene | **Y$_3$Fe$_5$O$_{12}$** | Ferrimagnet insulator | 27 meV | Exp, transport [56] |
| graphene | **EuO** | Ferromagnet insulator | 84 meV (electron) 48 meV (Holes) | Theory, DFT [39] |
| graphene | **MnO$_2$** | Ferromagnet insulator | 176 meV | Theory, DFT [81] |
| graphene | **CoFe$_2$O$_4$** | Ferrimagnet insulator | 45 meV (electron) 49 meV (Holes) | Theory, DFT [39] |
| graphene | **BaMnO$_3$** | Antiferromagnet semiconductor | 300 meV | Theory, DFT [83] |

*Note: We use the formula of Zeeman splitting energy $E_z = B_Z g \mu_B$, where $g$ is the Lande factor and $\mu_B$ is the Bohr magneton, to convert the exchange Zeeman field $B_Z$ to Zeeman splitting energy $E_z$. The "19.8-287.9 T" magnetic fields correspond to "2.29-33.37 meV" in the energy unit.

## 2.1 Graphene/BiFeO₃ heterostructure

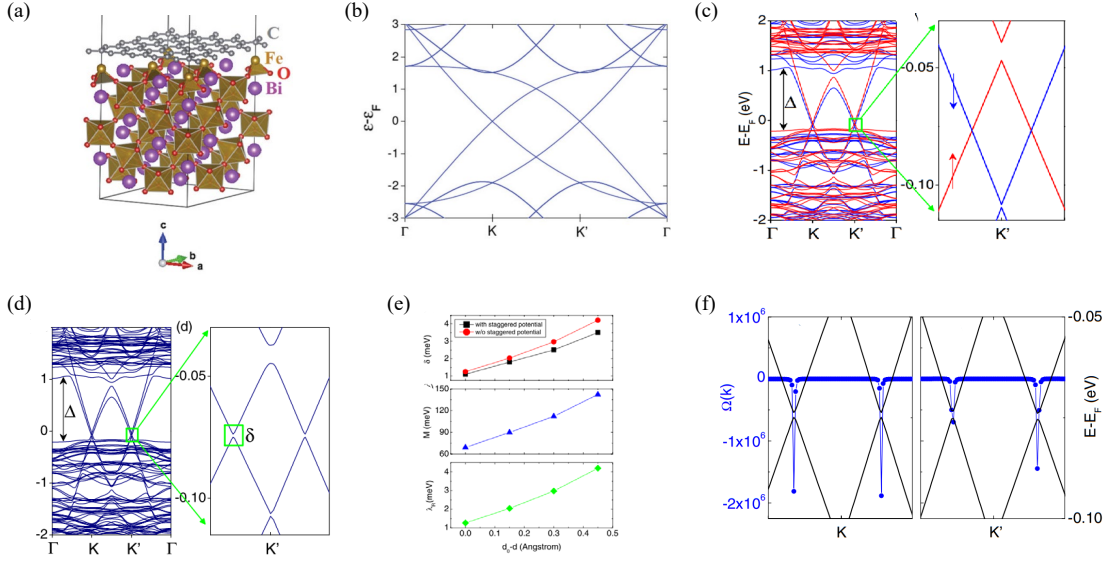

**Fig. 2. Quantum anomalous Hall effect of graphene on BFO.** (a) Supercell for graphene on BFO. (b) Band structure of pristine graphene. (c) Spin-resolved band structure of graphene on BFO. (d) Band structure with spin-orbit coupling considered. (e) Band gap, exchange field and Rashba spin-orbit coupling strength as a function of separation between graphene and BFO. (f) Non-zero Berry curvature along high symmetry lines. Figure adapted with permission from: ref [36], American Physical Society (a-f).

Graphene is a unique 2D material with a honeycomb lattice formed by carbon atoms, which has become an ideal prototype material for engineering the QAHE due to its linear Dirac band. However, graphene is not magnetic in its pristine form and also has extremely weak intrinsic spin-orbit coupling. Nevertheless, theoretical studies proposed that Chern insulating state in graphene could be realized by introducing both magnetic exchange field and Rashba SOC [44]. Therefore, introducing a long-range magnetic order in graphene is crucial in realizing the Chern insulating states. BiFeO₃ (BFO) has been considered as an appropriate substrate in a pioneer theoretical work [36] because of three reasons: First, BFO is insulating so that graphene is the only transport channel; second, the Fe atoms in BFO has large exchange energy [45], so the 3$d$ orbitals of Fe can hybridize with graphene strongly to induce sizable exchange and spin-orbit coupling. Third, the BFO is antiferromagnet, its dipolar magnetic field is weak and the signal of quantum anomalous Hall will not be disturbed by the normal quantum Hall effect.

BFO has a perovskite structure in which the magnetic plane is aligned, and neighboring planes have opposite spin polarizations [46]. Figure 2a shows the structure of graphene on an antiferromagnetic oxide BiFeO₃ (G/BFO) system. For pristine graphene, the band structure of isolated graphene is spin degenerated, and the Dirac points are located at K and K' points (Fig. 2b). When graphene is coupled to a BFO substrate, the π orbitals of graphene will be magnetized by the exchange interaction with the 3$d$ orbitals of the Fe layer in BFO, inducing a sizable exchange field. As shown in Fig. 2c, the spin-polarized band structure shows that the spin-polarized π bands at K point have an exchange splitting of $\lambda_{EX}$ ~70 meV. When the spin-orbit coupling is further considered, a small band gap opens at the crossing bands with opposite spin orientations (Fig. 2d). The exchange field and crossing band gap can be further enhanced by the separation between graphene and BFO. For example, when the separation is reduced by 0.5 Å, the strength of the exchange field and SOC can increase by a factor of 3 (Fig. 2e). The most important finding is that once summed over all the occupied valence bands, the Berry curvature Ω shows non-zero and has the same sign near the K and K' valley points (Fig. 2f). Consequently, its integration over the Brillouin zone leads to a nonzero Chern number. Because of the absence of an external magnetic field, this nonzero Chern number is the long-sought QAHE. As a pioneering theoretical prediction, the G/BFO work has inspired lots of theoretical and experimental works, including recent graphene/2D magnet Wan der walls heterostructures [47-50]. Compared with the 2D Wan der walls heterostructures, the graphene/magnetic oxide heterostructures usually experience a stronger exchange interaction due to the shorter separation of ~ 2.5 Å, while the former is usually at ~ 3.5 Å.

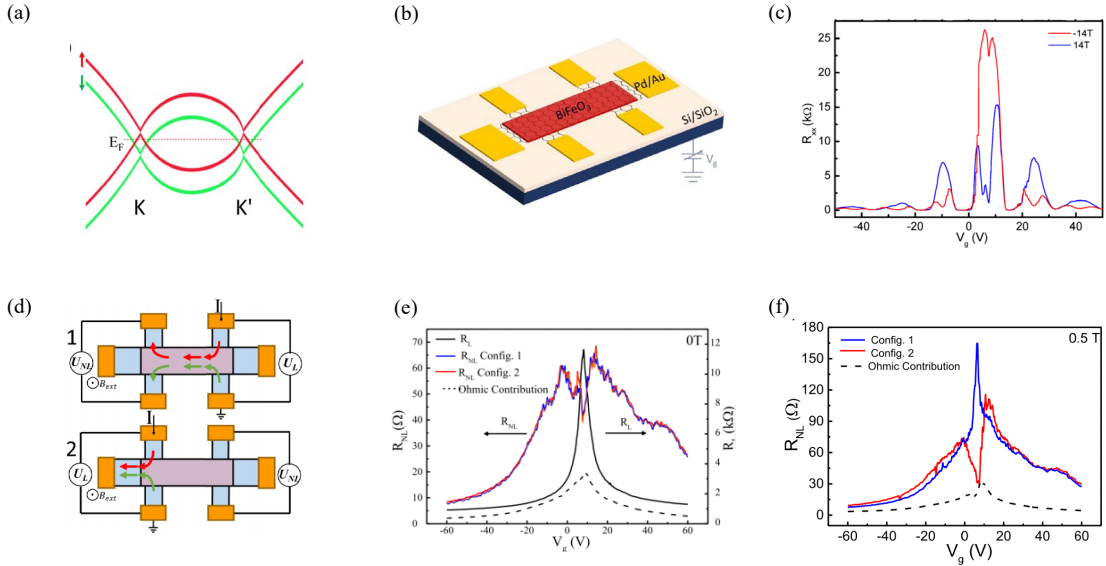

**Fig. 3. Zeeman spin Hall effect of graphene on BFO.** (a) Diagram for graphene band structure with Zeeman splitting. (b) The schematic diagram for the G/BFO Hall bar structure. (c) Asymmetric transport behavior under a reversed external perpendicular magnetic field of 14 T. (d) Schematics for non-local transport with two configurations. (e) and (f) Non-local resistivity measured under external magnetic fields of 0 and 0.5 T, respectively, with configurations 1 (blue) and 2 (red). Figure adapted with permission from: ref [51, 52], American Physical Society(a-f).

Experimentally, the magnetic exchange field of graphene on BFO can be revealed by the Zeeman spin Hall effect [51]. As illustrated in Fig. 3a, the spin-up and spin-down subbands are separated by the exchange field induced by Zeeman splitting, and the Dirac cones are lifted in opposite directions. Near the Dirac point, the graphene has both electron- and hole-like carriers but has opposite spins. Under Lorentz force, the electrons and holes will transport in opposite directions with a pure spin current, where the non-local voltage can be detected. Figure 3d shows two non-local configurations used to probe the spin-polarized signal. It can be seen that the two configurations show similar non-local signals without applying any magnetic field (Fig. 3e). Under a small magnetic field of 0.5 T, the non-local resistance of configuration 2 is quite close to the Ohmic contribution at the Dirac point. While for configuration 1, the non-local resistance is much stronger than the Ohmic contribution (Fig. 3f). The different non-local transport can be understood by the opposite Lorentz force for electrons and holes. For configuration 1, the electrons and holes diffuse to the two opposite edges due to Lorentz force induced by an external magnetic field, resulting in a voltage drop between the two voltage probes. For configuration 2, the electrons and holes driven by Lorentz force cannot reach the graphene area with the non-local probes. As a result, the non-local signals notably suppressed the Zeeman spin Hall effect. From the Zeeman spin Hall effect, the proximity-induced strength of the exchange field in graphene on BFO is 19.8 - 287.9 T [51].

Interfacial exchange coupling can also give rise to spin splitting of Landau levels (LLs), which can be studied by local transport [52] (Fig. 3b). As shown in Fig. 3c, the G/BFO hybrid device exhibits asymmetric transport behavior under reversed external perpendicular magnetic field. For negative field, interfacial exchange coupling gives rise to the spin splitting of $N \neq 0$ LLs and a quantum Hall metal state for $N = 0$ LL. While for positive magnetic fields, only $N = 0$ LLs are split, as there is a splitting resistance peak at the $N = 0$ LL (near the Dirac point). The splitting of nonzero LLs is attributed to the spin splitting but not valley splitting because the lifting of valley degeneracy in nonzero LLs is much more difficult than that in $N = 0$ LL. Meanwhile, the exchange field has a notable dependence on the external magnetic field, and it is much stronger under a negative magnetic field than that under a positive magnetic field. Moreover, using BFO as a top gate, the Zeeman splitting and energy gap of $N = 0$ LL can be further modulated by an electric field [53].

*2.2 Graphene/$Y_3Fe_5O_{12}$ heterostructure*

The exchange field by coupling graphene to a magnetic insulator can also be revealed by anomalous Hall effect (AHE). $Y_3Fe_5O_{12}$ (YIG) is garnet and possesses a cubic crystal structure. There are three sub-lattices in the garnet structure; the dodecahedra site occupied by the three Y ions, the octahedral site occupied by two Fe ions and the tetrahedral site occupied by three Fe ions. YIG film shows in-plane magnetic anisotropy, with a Curie temperature of 560 K [54]. By coupling graphene with YIG (G/YIG), magnetism can be induced in graphene but without sacrificing its excellent transport properties [55]. Figure 4a illustrates the spin-polarized transport of graphene on YIG (G/YIG). Due to the lift of spin degeneracy, the unbalanced spin-up and spin-down electrons will lead to spin-dependent quantum transport. The magnetic hysteresis loops show an in-plane magnetic anisotropy (Fig. 4b). The inset shows the AFM topographic image of a typical YIG film. The root means square surface roughness in 10-nm-thick YIG films can be as low as ~0.06 Å, which ensures an atomically sharp interface between graphene and YIG. The hybridization between $\pi$ orbitals in graphene and the nearby spin-polarized $d$ orbitals in YIG gives rise to the exchange interaction required for long-range ferromagnetic ordering. As shown in Fig. 4c, the G/YIG device shows a nonlinear Hall behavior at 2 K. The inset shows the nonlinear Hall data after subtracting the linear ordinary Hall background. At 2 K, the AHE of G/YIG has a magnitude of ~ 0.09 ($2e^2/h$).

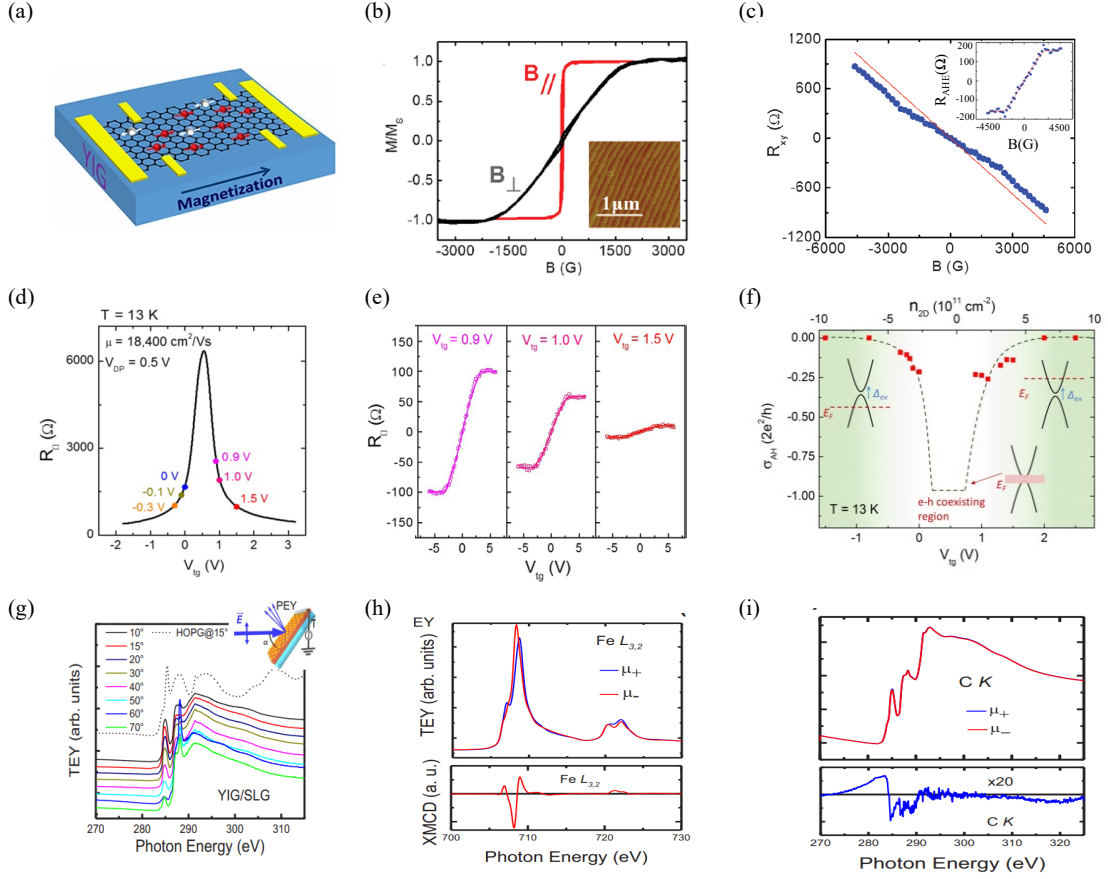

**Fig. 4. Anomalous Hall effect of graphene on YIG.** (a) Illustration of spin-transport in G/YIG heterostructure. (b) Magnetic hysteresis loops in perpendicular and in-plane magnetic fields. Inset is the AFM topographic image of YIG thin film surface. (c) The total Hall resistivity data at 2 K (anti-symmetrized) with a straight red line indicating the ordinary Hall background. The inset shows the nonlinear Hall resistivity after the linear background is removed from the data in (c). (d) Top gate voltage dependence of the sheet resistance of graphene sandwiched between YIG and hBN. (e) Anomalous Hall resistance in YIG/graphene/h-BN at different top gate voltages. (f) Top gate voltage and carrier density dependence of anomalous Hall conductance measured at 13 K. (g) Spectra acquired around the carbon K edge of the YIG/graphene, with the electric field vector of the linearly polarized beam pointing along the vertical direction, for several incidence angles α. (h) XAS spectra acquired around the Fe $L_{3,2}$ (left) and (i) C K (right) edges for both right- and left-handed circular polarization of the light. Figure adapted with permission from: ref [55-57], American Physical Society.

Apart from AHE, the observed nonlinear Hall can also be attributed to two other possible reasons. First, the coexistence of two types of carriers will also lead to nonlinear Hall in response to the external magnetic field [58]; second, the same Lorentz force related ordinary Hall effect but due to the stray magnetic field from the underlying YIG film [59]. By inserting $Al_2O_3$ between graphene and YIG, there is no measurable nonlinear Hall signal, indicating that the AHE in G/YIG is from the magnetic interaction between graphene and YIG. Moreover, the nonlinear Hall curves saturate at 2300 G, which is approximately correlated with the saturation of the YIG magnetization.

Thus the nonlinear Hall behavior is mainly attributed to the induced magnetism in graphene by YIG, as revealed by the AHE.

To improve the sample quality and further enhance the magnitude of AHE, the flat hBN is used as a top-gate dielectric [56]. Figure 4d shows the typical transfer curve of graphene on YIG using hBN as a top dielectric. The Dirac point locates at 0.5V, indicating a *p*-doping in G/YIG heterostructure. Closing to Dirac point at 0.9 V, the AHE is significantly enhanced and far away Dirac point at 1.5 V, the signal of AHE is decreasing (Fig. 4e). The dependence of AHE on carrier density is summarized in Fig. 4f. Near the Dirac point, the largest AHE can reach up to ~0.25 ($2e^2/h$). However, additional oscillatory features are observed when the Fermi level is further close to the Dirac point due to the multi-carriers in the e-h coexisting region. From the temperature dependence of AHE, the exchange coupling strength of G/YIG is ~27 meV [56].

The induced magnetism in G/YIG heterostructure can be directly detected by x-ray magnetic circular dichroism (XMCD) [57]. Figure 4g shows the spectra acquired near the carbon *K* edge, with the electric field vector of the linearly polarized beam pointing along the vertical direction. The changes of the XAS line shape show that the spectral feature around 285 eV, which is assigned to the C $1s \rightarrow \pi$ transition of $1s$ core electrons into unoccupied π states, is remarkably strong for grazing incidence angles and tends to vanish as the incidence angle is increased towards normal incidence. This strong angular dependence of the XAS line shape is due to different orbital orientations. It demonstrates that the π unoccupied states are aligned out-of-plane, as is expected for the C pz orbitals in the graphene-type layered sp$^2$ coordination. Figure 4h, i shows XAS spectra acquired around the Fe $L_{3,2}$ (left) and C K (right) edges. The Fe $L_{3,2}$ XAS and XMCD spectrum, which shows positive and negative peaks corresponding to tetrahedral sites (Tet $Fe^{3+}$) and octahedral (Oct $Fe^{3+}$), respectively, which have antiparallel coupled spins [60]. The most important finding is the observation of the obvious XMCD contrast at the C *K* absorption edge. The C *K* spectrum has the maximum negative asymmetry at 285-290 eV, indicating an induced magnetism in graphene on YIG and the magnetism mainly comes from the spin-polarized C $2P_Z$ orbitals [61].

Apart from AHE, the magnetic exchange field of G/YIG can also be studied by spin transport. Figure 5a shows the geometry of spin transport in graphene on YIG [62]. To measure the spin current, a current source is applied between E2 and E1, where E2 is a ferromagnetic metal, which serves as a spin injector because of the spin-dependent density. Once the spin is injected, the spins in graphene under E2 will diffuse in both directions toward E1 (a spin current with charge current) and E3 (a spin current but without charge current). Then the pure spin current can be detected by the voltage across E3 and E4, where E3 is the spin detector. Depending on the parallel or antiparallel between E2 and E3, the measured voltage can be modulated. By swapping an in-plane field, there is a sharp drop in non-local resistance at B = 35 mT, which can be attributed to the switching of magnetization of E2 and E3 into the opposite direction. To further modulate the spin current, a small field of 15 mT is applied, which is strong enough to saturate the YIG magnetization but smaller than what is required for switching electrodes (35 mT). As shown in Fig. 5c, when the magnetic field is rotated in the graphene plane, the non-local resistance has a periodical change. For parallel configuration (blue), the non-local MR is minimum due to the dephasing of the injected spins in graphene. While for antiparallel configuration (red), the injected spins are maximum. These experiments demonstrate that spin current can be modulated by the proximity exchange field between graphene and YIG.

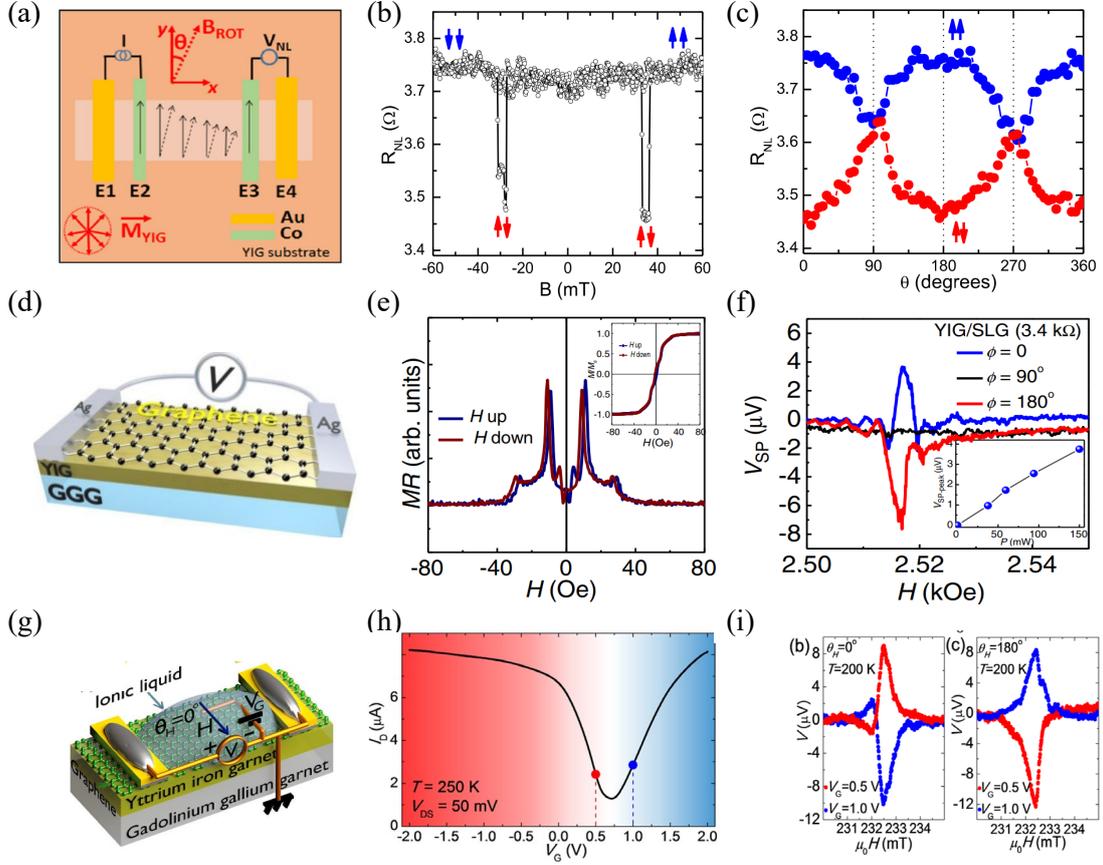

**Fig. 5. Spin transport of graphene on YIG.** (a) Schematic of the non-local spin transport measurement geometry. E2, E3 are magnetic electrodes and E1, E4 are non-magnetic electrodes. The arrow indicates the magnetization direction of YIG. (b) Non-local MR curves as a function of an in-plane magnetic field. (c) Non-local MR as a function of the direction of the magnetic field. A fixed B = 15 mT is applied in the YIG plane. (d) Schematic of the graphene/YIG heterostructure for spin pumping measurement. (e) Local MR as a function of out of the magnetic field. The inset shows the magnetization of YIG with the field. (f) Field scan spin-pumping dc voltage measured with 9.4 GHz microwave driving with power 150 mW in graphene. (g) Schematic of spin pumping measurement using ionic liquid gating. (h) Transfer curve of G/YIG heterostructure at 240 K. The red is the hole carrier, and the blue is the electron carrier. (i) Generated voltage by the FMR and application of the gate voltage Vg = 0.5 V (red lines) and Vg = 1.0 V (blue lines). Figure adapted with permission from: ref [62, 65, 68], American Physical Society.

One key for spintronics is the conversion of signals carried by spin into charge signal, which involves the spin-pumping effect and the inverse spin Hall effect [63, 64]. The interfacial interaction between graphene and YIG can be used to spin pumping from a YIG into graphene layers and observed by the electric detection of microwave-driven ferromagnetic resonance (FMR) in graphene [65]. Figure 5d shows the setup for the local MR and spin pumping measurement. Applying a forward and backward in-plane field, the local MR shows two identical peaks with small hysteresis, which correlate with the magnetization of YIG, suggesting the magnetic exchange interaction by the MPE. The spin-current to charge-current conversion can be studied by FMR measurement. The nanovoltage for the spin-pumping voltage was obtained with the field sweep. Figure 5f shows a large peak at the FMR position, which changes sign when the sample is rotated by 180° and falls to noise when the angle is 90°. Moreover, there is an asymmetry between the positive and negative peaks, similar to the case of Py/YIG [66], which indicates ferromagnetic order in G/YIG due to the proximity effect. The spin-current to charge-current effect is attributed to the inverse Rashba-Edelstein effect by extrinsic spin-orbit coupling in graphene [67].

By using liquid gating, the spin pumping can be further tuned by gate voltage [68], as illustrated in Fig. 5g. The ionic liquid is controlled by an electric field, leading to electron and hole doping of graphene. Figure 5h shows the dependence of drain current on the gate voltage. The minimum of the $I_D$ at 0.8V represents the Dirac point. The left side is the hole carrier, and the right side is the electron carrier. Figure 5i shows the detected voltage under FMR conditions. Since the generated spin current, which is converted from the angular moment of YIG to SLG, is in proportion to the area swept by the YIG magnetization during the precession. Thus, the spin current has a peak at the FMR field, where the microwave power absorption and precession angle are the largest. When the carrier type is changed by gate voltage, the change in the carrier type is accompanied by the switching of the voltage sign for both the 0° and 180° directions of the field. As a result, the sign of the spin-charge current conversion can be tuned by the gate voltage. Moreover, the spin-charge conversion current is independent of the applied top gate voltage, which can not be attributed to the Rashba-Edelstein effect. Instead, the intrinsic SOC is dominant over the Rashba-like SOC in graphene. The inverse spin Hall effect is the dominant spin-charge conversion so that the polarity of the spin-charge conversion current can be switched by the gate voltage.

*2.3 Graphene/EuO heterostructure*

EuO is a ferromagnetic oxide with a Curie temperature of 69 K and a small band gap (1.1eV). Also, it has potential applications in spintronics devices [69]. By coupling graphene with EuO (G/EuO), various applications have been obtained, such as single-electron transistor [70], highly tunable non-local transistor [71] and large anisotropic magnetoresistance [72]. Figure 6a shows the crystalline structure of G/EuO heterostructure [39]. Based on this structure, the induced magnetism of graphene in the G/EuO heterostructure is theoretically explored, and it is found that the magnetism can be induced in graphene through the exchange interactions between graphene and magnetic atom Eu. Exchange interactions produce exchange splittings, which can induce spin splitting. The exchange interaction between EuO and graphene can be described as a Zeeman splitting $\Delta \approx cJ \langle S_z \rangle$, in which $J$ is the spatial average of the exchange integral. Since the atomic wave function of EuO/Al and EuO/G are similar, it is assumed that $J$ is treated the same in two systems. Based on the J value of EuO/Al, the exchange splitting of G/EuO heterostructure is estimated to be 5 meV [37].

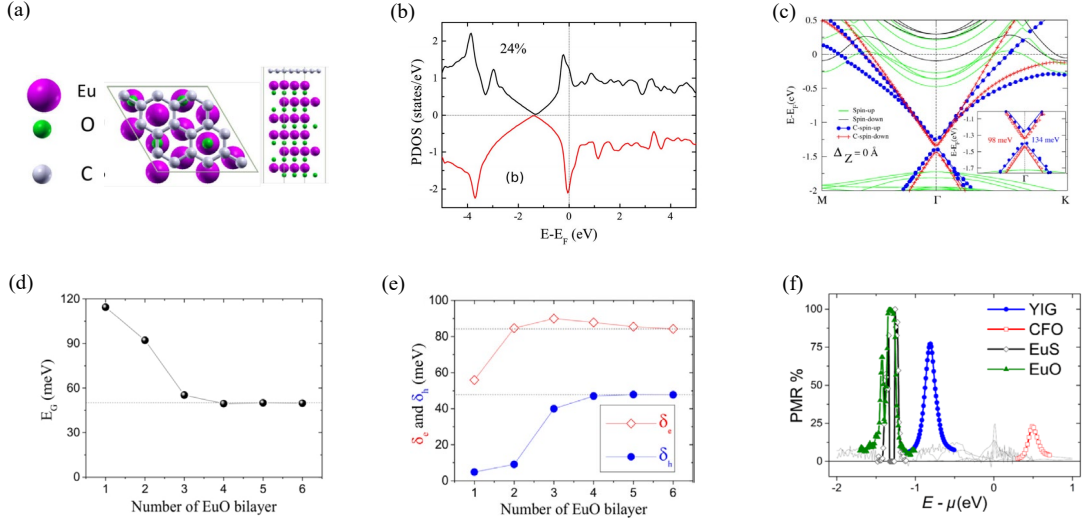

**Fig. 6. Spin polarization and PMR of graphene on EuO.** (a) Top and side views of the G/EuO heterostructure. (b) The total density of states of P$_z$ orbitals of graphene. (c) Spin-resolved band structure of G/EuO heterostructure. (d) Thickness-dependent Dirac cone band gap in G/EuO. (e) Thickness-dependent spin-splitting of the electron and hole bands at the Dirac cone δ$_e$ and δ$_h$ in G/EuO. (f) Proximity magnetoresistance as a function of energy with respect to the Fermi level for YIG (blue circles), CFO (red squares), EuS (black diamonds), and EuO (green triangles) using temperature smeared conductances at T = 300, 300, 16, and 70 K, respectively. Figure adapted with permission from: ref [39, 73,74], IOP Publishing, American Physical Society.

Exchange interaction will induce the spin polarization in graphene, and the average spin polarization $p$ is about 24% from the total density of states (Fig. 6b), where $p = (n^↓−n^↑)/(n^↓+n^↑)$ [73]. The exchange interactions between graphene and EuO also open a band gap in graphene. In the energy-optimized structure, the difference between the spin-up and the spin-down band gap is 36 meV (Fig. 6c). Moreover, the spin-splitting energy can be tuned by the distance between EuO and graphene. As the distance between EuO and graphene decreases, the overlap between C $p_z$ and Eu 4$f$ orbitals will be enhanced, and more electrons are transferred to the graphene. As a result, the spin splitting of graphene will gradually increase as the distance between EuO and graphene decreases [73]. Furthermore, the thickness of EuO will affect the exchange interaction in proximity with EuO. With the increase of EuO thickness, the band gap of graphene and the electron and hole bands at the Dirac cone tend to reach the bulk values (Fig. 6d, e) [39]. In addition, there is a large proximity magnetoresistance (PMR, up to100%) in the G/EuO heterostructure (Fig. 6f). [74]. Moreover, the MPE in G/EuO will not only lead to the spin splitting of graphene but also introduce intervalley interactions leading to the nonlinear distribution of Γ points [38].

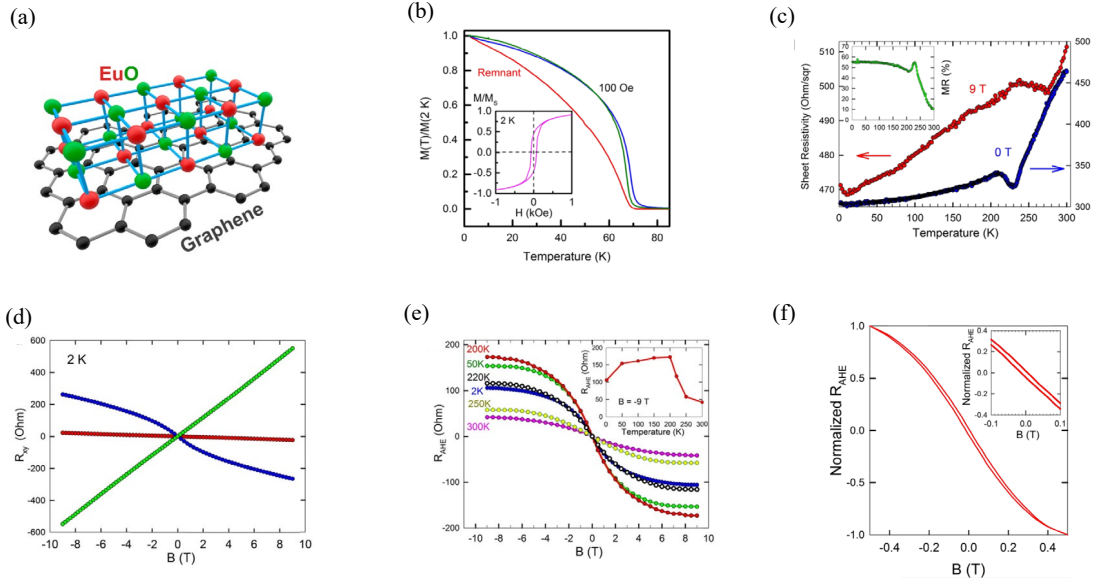

**Fig. 7. Anomalous Hall effect of graphene on EuO.** (a) Ball-and-stick model of G/EuO heterostructure. (b) Temperature dependence of the normalized magnetization of the films in a magnetic field 100 Oe, the EuO/graphene structure (blue) and the EuO/graphene(+Eu) structure (green). The inset: magnetic field dependence of normalized magnetization of the EuO/graphene(+Eu) film at 2 K. (c) the EuO/graphene structure in zero magnetic field (blue) and in a magnetic field of 9 T (red, notice different scales); inset: MR of the EuO/graphene structure. (d) Magnetic field dependence of the transverse resistance of graphene (green), EuO/graphene (blue), and EuO/graphene (+Eu) (red). (e) Magnetic field dependence of the AHE resistance in the EuO/graphene structure at 2 K (blue), 50 K (green), 200 K (red), 220 K (black), 250 K (yellow), and 300 K (magenta). The inset: temperature dependence of the saturated (taken at −9 T) AHE resistance in the EuO/graphene structure. (f) The hysteresis of the EuO/graphene transverse resistance on the magnetic field dependence at 2 K. The inset is an enlargement of the magnetic field area. Figure adapted with permission from: ref [76], American Chemical Society.

Experimentally, high-quality crystalline EuO can be directly grown on exfoliated graphene. More importantly, after epitaxial growth of the EuO layer, the graphene can still maintain high quality [75]. The effect of EuO magnetic film on the magnetic transport properties of graphene has been studied [76]. Molecular beam epitaxy was used to synthesize EuO films on CVD graphene to explore the magnetic properties of graphene. The magnetic transition temperature in G/EuO heterostructure is 220 K (Fig. 7c), well above the Curie temperature 69 K of EuO (Fig. 7b). The reason is the extra spin-orbit coupling effect caused by the heavy atom effect of Eu atoms [77]. To further study the proximity-induced magnetism in graphene, the transverse resistance is measured in Fig. 7d. At T = 2 K, the relationship between transverse resistance ($R_{xy}$) and magnetic field (B) is linear in graphene and EuO/graphene (+Eu). However, for EuO/graphene structure, the $R_{xy}(B)$ presents a nonlinear behavior. To determine whether this nonlinear Hall is due to the magnetism of graphene, AHE resistance RAHE(B) at different temperatures has been measured. As shown in Fig. 7e, the nonlinearity reduces when the temperature exceeds Curie temperature $T_c$(220 K). Also, AHE resistance ($R_{AHE}$) tends to saturate below 220 K. However, the saturation RAHE drops sharply when the temperature exceeds 220 K, which is consistent with the Curie temperature of graphene (Fig. 7c). It can be inferred that there is a ferromagnetic transition at the transition temperature. Moreover, the induced magnetism in graphene is also supported by the hysteresis of the transverse resistance (Fig. 7f). Thus, the nonlinear Hall behavior is related to the induced magnetism in graphene. These experimental results prove that the nonlinear Hall in G/EuO heterostructure is the anomalous Hall effect.

*2.4 Graphene/CoFe$_2$O$_4$ heterostructure*

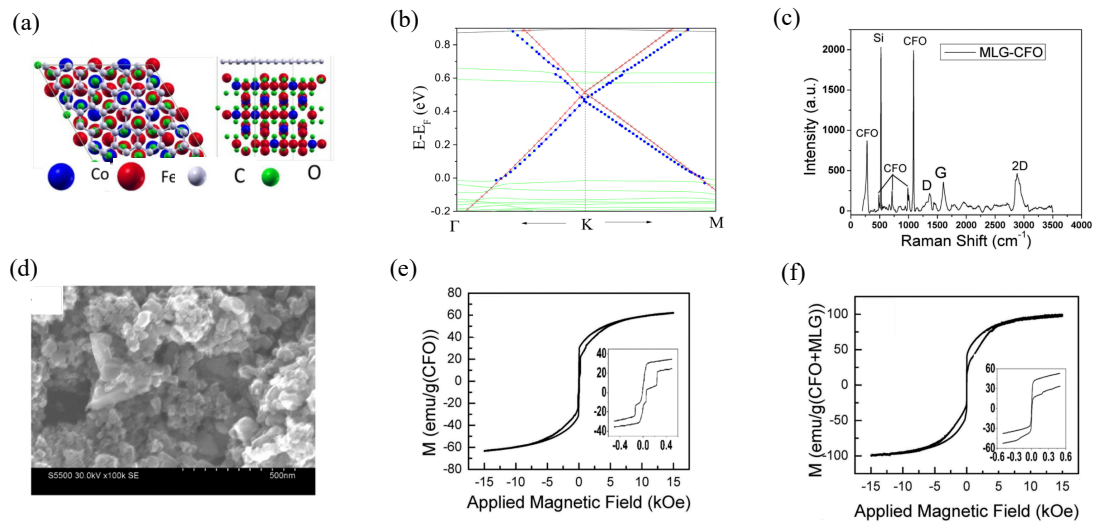

**Fig. 8. Hysteresis loop of graphene on CoFe$_2$O$_4$.** (a) Side view and top view of CoFe$_2$O$_4$/graphene ball and stick model. (b) Band structure of CoFe$_2$O$_4$/graphene. (c) Raman Spectra of CFO and graphene composite. (d) SEM of CFO and graphene composite. (e) Hysteresis loop of CFO. (f) Hysteresis loop of the CoFe$_2$O$_4$/graphene composite. The insets in both graphs show the low field portion of the hysteresis loops. Figure adapted with permission from: ref [39, 78], IOP Publishing, AIP Publishing.

CoFe$_2$O$_4$ is a ferromagnetic insulator oxide, containing both Co and Fe as magnetic atoms. The interaction graphene with CoFe$_2$O$_4$ (G/CFO) can also induce magnetism in graphene. The optimal G/CFO heterostructure has been calculated (Fig. 8a) [39]. Based on this configuration, the band structure in the G/CFO heterostructure was calculated (Fig. 8b). The exchange splitting for electrons and holes are 45 meV and 49 meV, respectively. Due to the exchange splitting, spin polarization is generated in graphene, resulting in a spin-dependent gap. The spin-up and spin-down gaps are 12 meV and 8 meV, respectively [39]. In addition, the composites of multilayer graphene and CFO had been explored. Raman and SEM confirm the successful synthesis of composite and uniform distribution of CFO on multilayer graphene (Fig. 8c, d). Further experiments show that the magnetic properties of G/CFO composites were greater than that of CFO, indicating that the induced magnetism in graphene is due to the proximity effect of CFO (Fig. 8e, f) [78].

## 2.5 Graphene/MnO₂ heterostructure

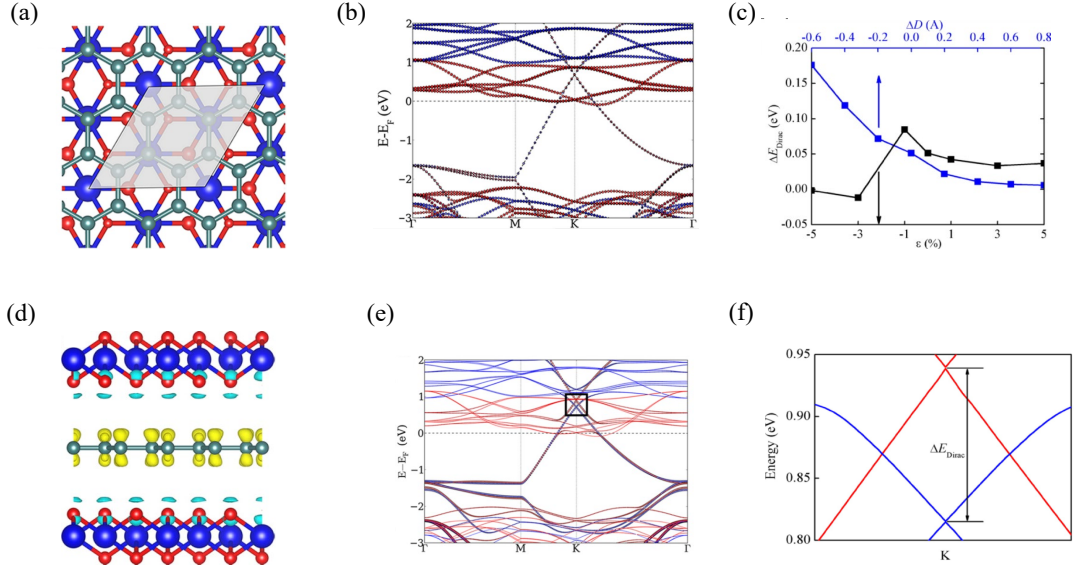

**Fig. 9. Spin splitting of graphene on MnO₂.** (a) Top view of Mn-centered configurations of the MnO₂/graphene. The grey, blue, large red, and small red balls represent C, Mn, O-first, and O-second atoms, respectively. (b) Spin-polarized band structure of the MnO₂/graphene. (c) $\Delta E_{Dirac}$ as a function of the interlayer distance (blue) and strain (black). (d) Charge density difference for the MnO₂/graphene/MnO₂ system. The yellow and cyan regions represent charge depletion and accumulation, respectively. (e) Spin-polarized band structure of the MnO₂/graphene/MnO₂. (f) Band structure of graphene in MnO₂/graphene/MnO₂. Note: Red is the spin majority, and blue is the spin minority. Figure adapted with permission from: ref [78], American Chemical Society.

MnO₂ is a ferromagnetic semiconductor with a high Curie temperature which contains Mn as magnetic atoms [79]. Also, G/MnO₂ has been obtained experimentally, which can be used to explore the interaction between graphene and MnO₂ [80]. Theoretically, the electronic properties of the interface between MnO₂ and graphene have been studied [81]. The configurations of G/MnO₂ heterostructure and band structure of Mn-centered MnO₂/graphene had been studied at the equilibrium position (the interlayer distance and binding energy is 3.03Å and 90meV) (Fig. 9a, b). The features of graphene and MnO₂ are well preserved, but due to the influence of graphene on MnO₂, half-metallic MnO₂ can be obtained, and spin polarization can be seen in graphene because of the proximity coupling to MnO₂. To further explore the role of graphene and MnO₂ in the interface, another sandwiched MnO₂/G/MnO₂ system has been studied. As shown in Fig. 9e, there is a clear spin splitting band in MnO₂/G/MnO₂ heterostructure and the spin splitting energy at the Dirac point is 176 meV (Fig. 9f). Here note that the induced magnetism in graphene is due to the interaction between the π orbitals of graphene and the O *2p* orbitals (antiferromagnetically aligned to Mn) [81]. Also, in Fig. 9d, there is another interesting phenomenon in this system, the charge dissipation has different effects on the two sublattices of graphene. As a result, the band gap of graphene will be open 3meV at Dirac point, which can be used for device fabrication. MPE relies on the exchange interaction and the orbitals hybridization at the interface, so it can be adjusted by applying strain and changing the distance between MnO$_2$ and graphene. As shown in Fig. 9c, applying tensile stress will weaken this proximity effect and reduce spin polarization; applying compression will weaken the exchange interaction on the one hand and increase the orbitals overlap on the other. Under the competition of these two mechanisms, the spin polarization reaches the maximum and minimum at ε = −1 and −3%, respectively. The distance between MnO$_2$ and graphene further directly affects the spin splitting energy.

*2.6 Graphene/BaMnO$_3$ heterostructure*

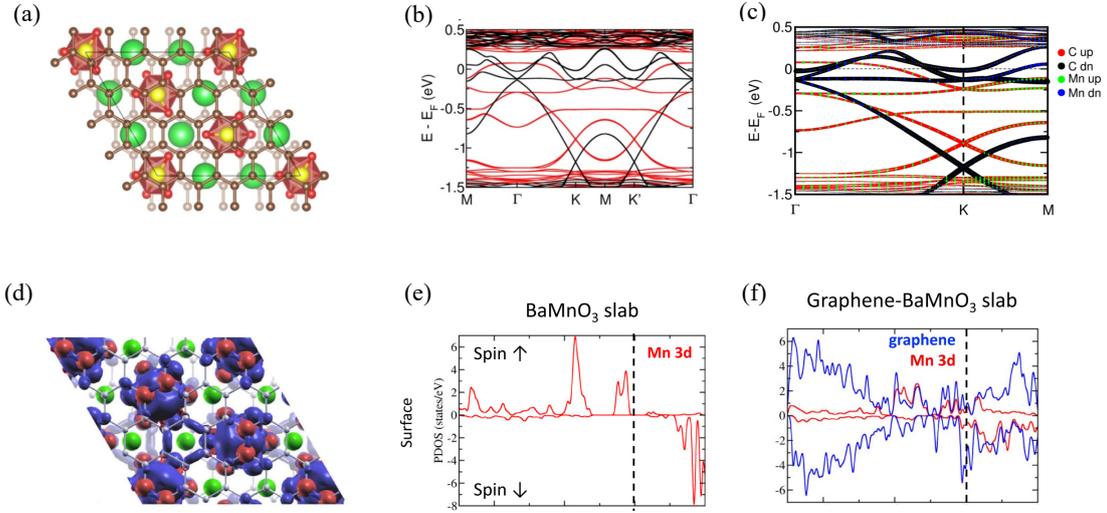

**Fig. 10. Mutual magnetic interaction of graphene on BaMnO$_3$.** (a) Ball and stick model of graphene/BaMnO$_3$ (Mn yellow, Ba green, O red, C gold, B light blue). (b) band structure of graphene/BaMnO$_3$. (c) Electronic band structure of graphene-BaMnO$_3$ projected on the C and surface Mn atoms for collinear spins. The density of states of surface layers in BaMnO$_3$. (d) Top view of the spin density (ρ ↑ − ρ ↓). Blue and red correspond to positive and negative isodensity, respectively. (e) Projected density of states for the BaMnO$_3$ and (f) graphene-BaMnO$_3$ slabs. Figure adapted with permission from: ref [83, 84], Springer Nature, American Physical Society.

Hexagonal 2H-BaMnO$_3$ is a representative type of magnetoelectric material and has a lattice constant that matches that of graphene [82]. The MPE in graphene/BaMnO$_3$ (G/BMO) heterostructure had been studied by DFT calculations [83]. The most stable structure of G/BMO heterostructure has been shown in Fig. 10a. Every Mn atom lies in the middle of the C-C bond. The vertical distance is 1.84Å. The band structure of graphene/BaMnO$_3$ is spin-polarized and shows a quasi-half-metal character (Fig. 10b). In addition, the hybridization of the C $sp^2$, Mn $3d$ and O $2p$ is found. The magnetism of the system is due to the excess charge transferred to graphene, which induces spin polarization in carbon (Fig. 10d). At the same time, the relative orientation of the two substances will result in a quasi-half-metal or magnetic semiconductor [83]. For G/BMO structure, on one hand, the interaction between graphene and BaMnO$_3$ will induce spin-polarization in graphene (Fig. 10b), on the other hand, graphene can also affect the magnetic properties of BaMnO$_3$, since several layers on the surface of BaMnO$_3$ change from antiferromagnetic to ferromagnetic (Fig. 10e, f) [81].

*2.7 Graphene/La$_{0.7}$Sr$_{0.3}$MnO$_3$ heterostructure*

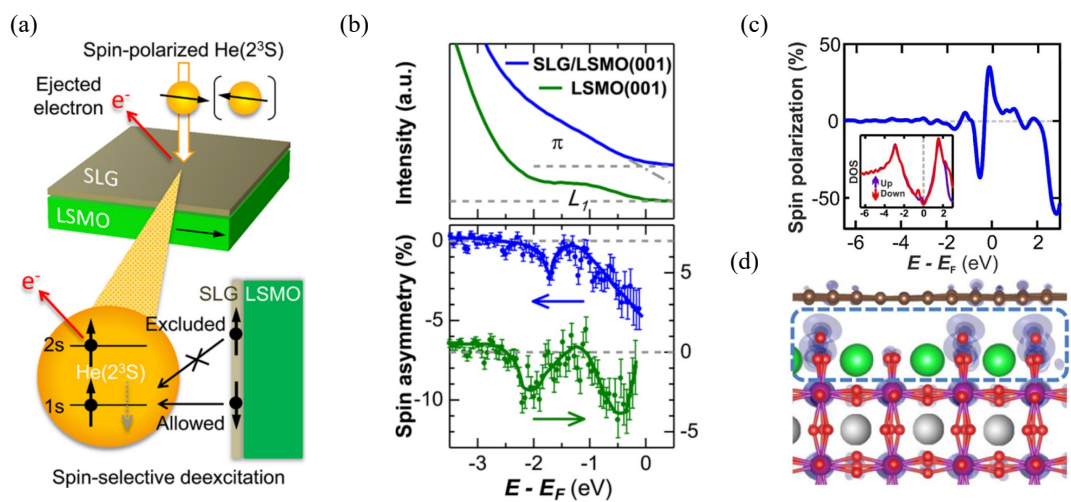

**Fig. 11. Direct evidence for the magnetic proximity effect of LSMO on graphene.** (a) Schematic illustration of the SPMDS measurements for the graphene/LSMO. (b) SPMDS and spin asymmetry spectra of graphene/LSMO (blue) and LSMO (green) after 1000 K annealing as a function of the binding energy E-E$_F$ in the energy region near the Fermi level. (c) Spin polarization of graphene/LSMO, inset is the density of states. (d) Side view of graphene/LSMO with different spin densities from -0.5 to 0 eV. Figure adapted with permission from: ref [85], American Chemical Society.

Spin-polarized metastable atom deexcitation spectra (SPMDS) have been used as a probing method to study MPE between graphene and La$_{0.7}$Sr$_{0.3}$MnO$_3$ (G/LSMO) [85]. SPMDS is very sensitive to local electronic structure on the surface of G/LSMO by measuring the intensity and energy distribution of ejected electrons by changing the spin direction of He(2$^3$S) (Fig. 11a). This technique can be used to detect the spin polarization state of graphene induced by LSMO. The linear features of SPMDS spectral in Fig. 11b demonstrate no noticeable change in π band structures of graphene around the Fermi level. Also, as for spin asymmetry signs, G/LSMO and LSMO are identical in the region close to E$_F$, indicating that the graphene π band is spin-polarized parallel to LSMO with the positive spin polarization. DFT has been used to calculate the spin-dependent density of states to further reveal the spin polarization in graphene induced by LSMO (Fig. 11c). Figure 11d shows that the spin polarization in graphene is indirectly caused by MnO in LSMO through the oxygen atom between Mn and graphene. Therefore, the MPE of LSMO on graphene is proved according to the spin asymmetry of the spin-polarized metastable atom deexcitation spectrum.

## 3. MPE at transition metal dichalcogenide and magnetic oxide insulator

Apart from the graphene-oxides magnetic proximity effect, the MPE between transition metal dichalcogenides (TMDs) and magnetic oxides has also become a hot topic recently. Compared with graphene, TMDs have various advantages in the application of spintronics. First, TMDs have band gaps due to their semiconductor properties, making it possible to use them in new field-effect transistors and optoelectronic devices [86]. Secondly, TMDs also provide an excellent platform to study the spintronic devices and its application with strong spin-orbit coupling two-dimensional materials [87]. Third, the single-layer TMD materials induce the valley Hall effect due to its broken inversion symmetry [88]. This makes practical applications in the newly formed field "valleytronics" possible. When applying an in-plane electric field, the carriers in different valleys flow to the opposite lateral edge. The broken inversion symmetry will also contribute to the K point valley-related optical selection for inter-band transitions [89]. This effect is of great significance in the design and application of valley-polarized devices. Additionally, TMDs are easily exfoliated and physically stable [90]. Due to TMD materials' unique properties, emergent phenomena can be induced in TMDs by coupling with a magnetic substrate. In this section, we discuss the interfacial MPE of TMDs and magnetic oxide insulators.

*3.1 $MoS_2$ /$BiFeO_3$ heterostructure*

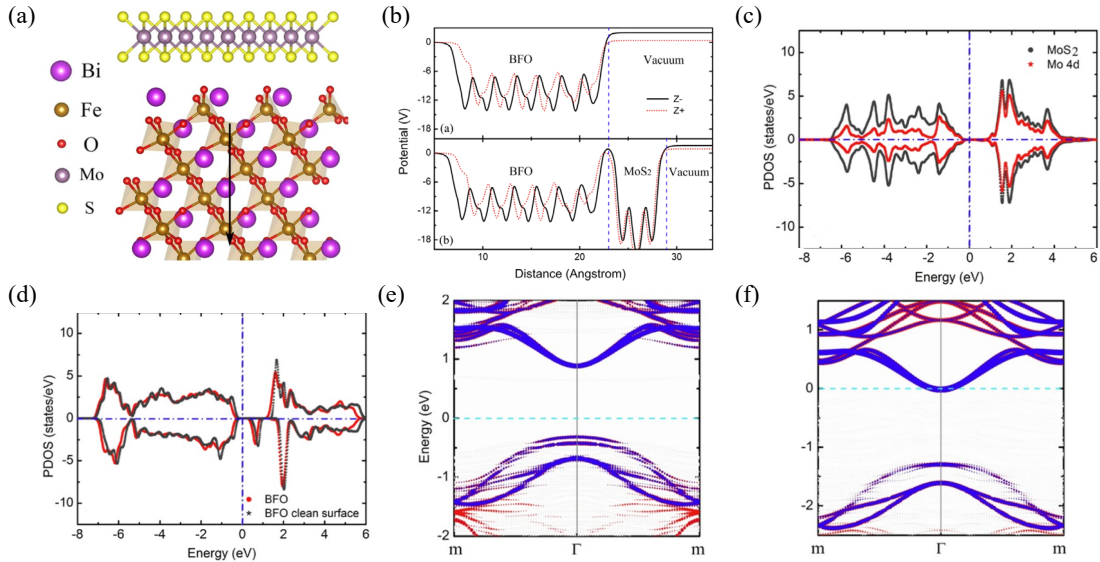

**Fig. 12. Large band offset of $MoS_2$ on BFO.** (a) Schematic image of a monolayer $MoS_2$/$BiFeO_3$ heterostructure (b) Planar-averaged electrostatic potential for the bare BFO(0001) surface (Top) and the $MoS_2$/BFO hybrid system (Bottom). The solid black line indicates the BFO Z− surface, whereas the red dotted line represents the BFO Z+ surface. We assumed that the backside of the BFO slab is grounded and shows zero potential. (c) Projected density of state (DOS) on the $MoS_2$, Mo 4d, and BFO surface

(one Fe-$O_3$-Bi trilayer plus the surface termination) for the MoS$_2$/BFO(0001) Z− system. (d) DOS of the clean BFO(0001) surfaces. (e) Fat band of the MoS$_2$ monolayer (red circles) and Mo 4d states (blue circles) for the MoS$_2$/BFO Z− system. (f) Fat band of the MoS$_2$/BFO for Z+ hybrid system. Figure adapted with permission from: ref [91], American Chemical Society.

The lattice constant of MoS$_2$ is $a$ = 0.3161nm. Its space group is *P63/mmc*. MoS$_2$ has a layered structure similar to graphene. Different layers are connected by van der Waal forces between sulfur ions, so the bulk MoS$_2$ is easily peeled off to obtain single layers. MoS$_2$ is chemically stable, with band gap 1.23eV (bulk) or 1.8eV (single layer) [92]. Compared to graphene, MoS$_2$ has stronger optical selectivity due to its broken inversion symmetry. The K, K' boundary has greater momentum space separation, so we can use SR-PLs to characterize the spin polarization induced by MPE [89]. By introducing a magnetic field from magnetic oxide, we can achieve MPE and induce spin polarization in MoS$_2$.

The MoS$_2$/BFO heterostructure was studied for its large band offset on oppositely polarized BiFeO$_3$(0001) [91]. Figure. 12a shows the heterostructure of MoS$_2$ on BFO (MoS$_2$/BFO). The black arrow presents the polarized direction. Z− represents the polarization direction along the arrow, while Z+ represents the polarization direction along the opposite direction of the arrow. The difference in polarization direction is due to the physical adsorption between MoS$_2$ and BFO when fabricating the heterostructure. Planar-averaged electrostatic potential of MoS$_2$/BFO system and bare BFO is shown as Fig. 12b. The electrostatic potential within the BFO demonstrates the existence of a depolarizing electric field that points opposite to the ferroelectric polarization direction [93, 94]. Besides, the electrostatic potential of the BFO Z− surface is higher by 1.59 V than that of the BFO Z+ surface. As for the MoS$_2$/BFO system, the absorption of the MoS$_2$ reduces the electrostatic potential difference to 0.64 V, suggesting that the MoS$_2$ monolayer took on the extra 0.95 V to compensate the electrostatic potential difference between the bare BFO Z− and Z+ surfaces. Projected DOS on the MoS$_2$, Mo 4*d*, and BFO surface for the MoS$_2$/BFO(0001) Z−system is presented by Fig. 12c, while the DOS of the clean BFO(0001) is shown in Fig. 12d. The PDOS of MoS$_2$ (Fig. 12c) shows a weak downward shift in energy, and the Fermi energy is located at the lower part of the band gap. In contrast, the energy position of the PDOS for the BFO Z− surface (Fig. 12d) seems to be unaffected by the MoS$_2$ adsorption. This result also demonstrates that the interaction between MoS$_2$ and the BFO(0001) surface is dominated by the physisorption mechanism. Compared to the freestanding MoS$_2$ monolayer, the band gap of the MoS$_2$ monolayer adsorbed on the BFO(0001) surface decreases to 1.2 eV. For the MoS$_2$/BFO(0001) hybrid system, the flat band of the MoS$_2$ monolayer and the Mo 4d orbitals are shown in Fig. 12e and Fig. 12f. As for the MoS$_2$/BFO Z- system (Fig. 12e), the Fermi energy lies 0.3 eV above the valence-band maximum of the MoS$_2$ monolayer, which retains the semiconducting characteristics. For the MoS$_2$/BFO Z+ system (Fig. 12f), the MoS$_2$ monolayer shows an n-type conducting behavior since the Fermi level has just crossed the conduction-band minimum. From the above analysis, researchers predict a large band offset of 0.9 eV occurring in the MoS$_2$ monolayer on the oppositely polarized BFO(0001) surfaces.

## 3.2 MoS$_2$/Y$_3$Fe$_5$O$_{12}$ heterostructure

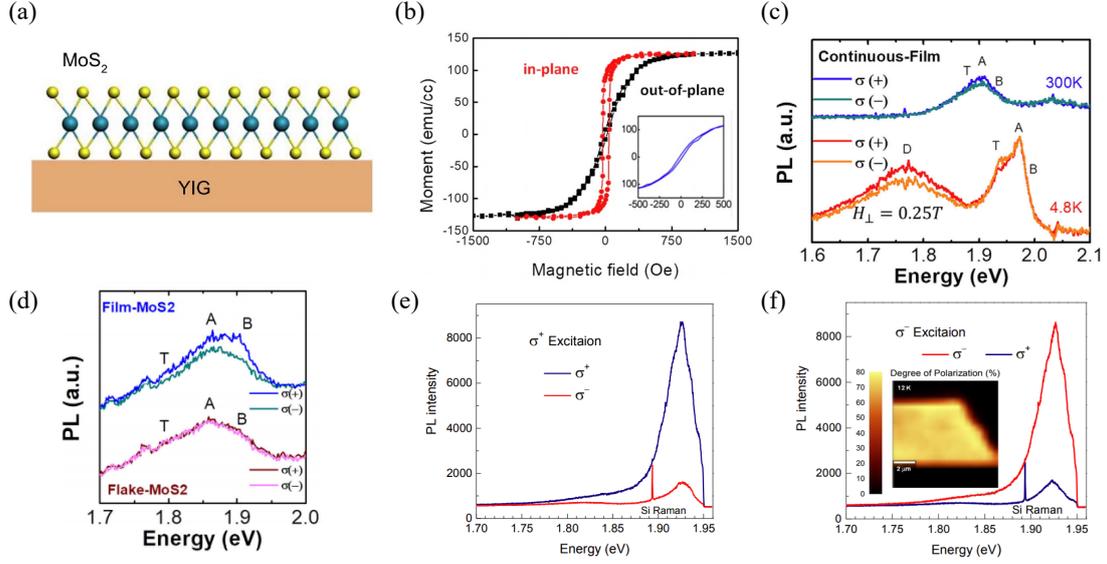

**Fig. 13. Spin polarization of MoS$_2$ on YIG.** (a) Schematic image of a monolayer MoS$_2$/YIG heterostructure. (b) Hysteresis loops obtained along the perpendicular (hard-axis, black curve) and in-plane (easy-axis, red curve) directions at 300 K. Inset: The center of the out-of-plane hysteresis curve exhibiting detectable coercivity. (c) SR-PL of film-MoS$_2$ obtained at 300 and 4.8 K under a perpendicular field of 0.25 T. A, T, and D correspond to the A-exciton, trion, and shallow state. (d) Room-temperature SR-PLs of film-MoS$_2$ and flake-MoS$_2$ on YIG film. (+) and σ(−) respectively represent PL components with left- and right-handed circular polarization. (e)(f) Polarization-resolved photoluminescence spectra of the trion by Right-hand (σ$^-$) and left-hand (σ$^+$) circular polarized excitation at 1.96 eV at 12 K, nearly on-resonance with the trion and A exciton. The peak at ~1.89 eV is assigned to the Si Raman feature, indicating that the system is in excellent calibration. The inset in Fig. 13f is a two-dimensional image of the degree of circular polarization η at 12 K, η = ~70%. Figure adapted with permission from: ref [95], John Wiley and Sons; ref [96], American Chemical Society.

YIG has short-range magnetic interaction with MoS$_2$. MoS$_2$ obtains anti-ferromagnetism induced by MPE at room temperature was researched in recent years [95]. Figure 13a shows the heterostructure of MoS$_2$/YIG. Figure 13b shows hysteresis loops obtained along perpendicular and in-plane directions at 300 K, which implies the in-plane anisotropic field can reach up to 3 kOe. In-plane hysteresis loop changes intensively when the magnetic field is around 17 Oe. The in-plane hysteresis effect is obvious, while the out-of-plane hysteresis effect is weak, indicating the existence of a minor perpendicular remanence at the zero field, despite the magnetism being dominated by in-plane shape anisotropy. Magnetic circular dichroism (MCD) is used to study the polarization of MoS$_2$/YIG. There is no MCD except at the point wherein MoS$_2$ was in direct contact with YIG (Fig. 13c), which reflects the interface-sensitive nature of MPE. Room-temperature SR-PLs of film-MoS$_2$ and flake-MoS$_2$ on YIG in Figure 13d shows that the intensity of left and right circular polarization is inconsistent in film-MoS$_2$ on YIG, which suggests the left- and right-handed circular polarization happens in film-MoS$_2$ on YIG and induces MPE. Moreover, there is no such phenomenon when placing flake-MoS₂ on YIG under the same condition. Therefore, we conclude that MPE only happens in single-layer MoS₂ on YIG and MPE is mainly induced by the interfacial effect between MoS₂ and YIG. Spin polarization also happens on MoS₂/YIG heterostructure under low temperatures [96]. Polarization-resolved photoluminescence spectra of the trion by right-hand ($\sigma^-$) and left-hand ($\sigma^+$) circular polarized excitation at 1.96 eV at 12 K (Fig. 13e and Fig. 13f) suggests that there is a big difference in the absorption of left and right circularly polarized light in a single layer of MoS₂ at low temperatures, which implies that spin polarization also happens at low temperature. Spin polarization induced by the YIG MPE of MoS₂ at low temperatures shows that under the excitation of circularly polarized light, MoS₂ achieves the time symmetry breaking, thus the valley Hall effect is produced, which illustrates the potential of MoS₂ as a valley spin device.

*3.3 MoS₂/CoFe₂O₄ heterostructure*

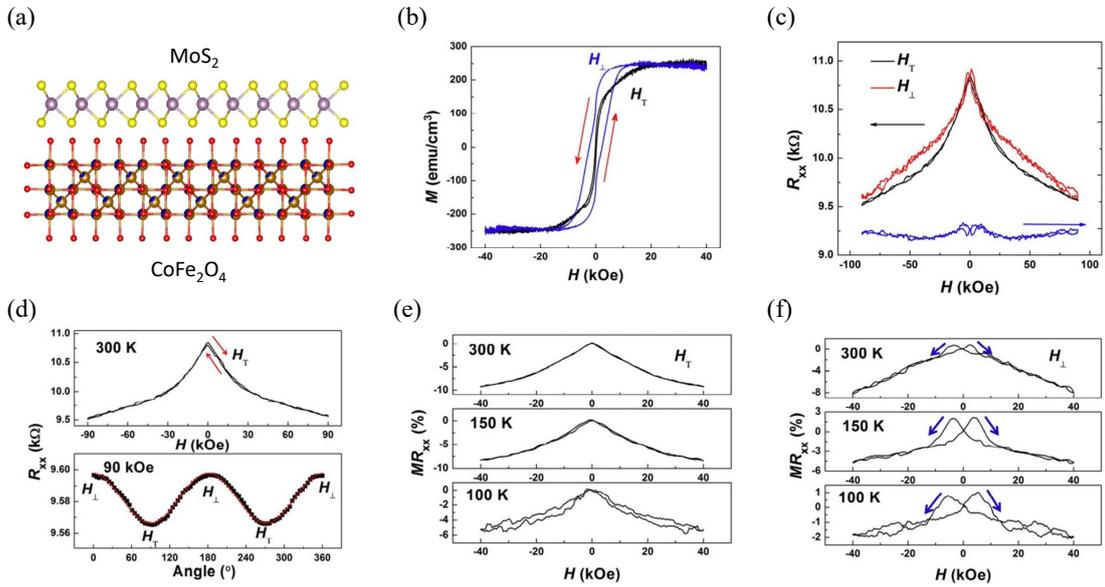

**Fig. 14. Room-temperature magnetoresistance of MoS₂ on CFO.** (a) Schematic image of a monolayer MoS₂/CoFe₂O₄ heterostructure. (b) Magnetic hysteresis loops of the CFO sample grown on single-crystal MgO substrate when the magnetic field is applied in the CFO plane transverse to the electric current (black loop) or perpendicular to the CFO plane (blue loop). (c) The difference of resistance of 1L MoS₂ when H is applied perpendicular to the MoS₂ plane or transverse to the current direction. (d) Top: The H-dependent resistance of monolayer MoS₂ when H is applied transverse to the electron current direction. Bottom: the angle-dependent resistance when H is fixed at 90 kOe. (e) The MR behavior of 1L MoS₂ sample when the H is applied in the direction of transverse to the electric current in the MoS₂ plane at 300, 150 and 100 K, respectively. (f) The MR behavior of 1L MoS₂ when the H is applied perpendicular to the MoS₂ plane at 300, 150 and 100 K, respectively. Figure adapted with permission from: ref [97], American Chemical Society.

The influence of MPE on magnetoresistance of MoS₂ at room temperature, that is, the magnetoresistance effect of MoS₂/CoFe₂O₄ at room temperature, is studied [97]. Figure 14a shows the schematic image of a monolayer MoS₂/CoFe₂O₄ heterostructure. Figure 14b shows the H dependence of the in-plane and out-of-plane magnetization (M) in CFO recorded at room temperature. The out-of-plane loop indicates a typical ferromagnetic behavior with a coercive field of about 2.3 kOe and a saturation field of about 12 kOe, while the in-plane magnetic measurement shows a very slim hysteresis loop with a small coercive field near to zero, suggesting the perpendicular anisotropy in the PLD-deposited CFO thin films. As Fig. 14c presents, the difference of resistance of single-layer $MoS_2$ when H is applied perpendicular to the $MoS_2$ plane or transverse to the current direction at room temperature under 90kOe magnetic field shows that resistance of $MoS_2$/CFO can rise to 12.7%, but magnetoresistance observed in the nano-ferromagnetic and antiferromagnetic insulators is usually less than 1%. The phenomena suggest that when a single layer of $MoS_2$ is adjacent to the ferromagnetic oxide CFO and external field is applied, spin accumulation will form on their contact interface, which in turn induces a strong magnetoresistance effect at room temperature [87, 98]. According to Figure. 14d, $R_{xx}$ is changed as a function of the sweeping angle (θ) when H is fixed at 90 kOe. The $R_{xx}$ oscillation follows a square of cosine relationship with the sweeping angle, as demonstrated by the fitted red curve. The amplitude of the resistance oscillation with the field angle looks like a signature of the spin Hall magnetoresistance (SMR), which indicates that the researcher obtained the H-intensity dependent MR and the H-direction dependent SMR in the $MoS_2$/CFO heterostructure. Additionally, The MR behavior of 1L $MoS_2$ sample when the H is applied in the direction of parallel (Fig. 14e) and transverse (Fig. 14f) to the electric current in the $MoS_2$ plane at 300, 150 and 100 K implies that the MR behavior along two different directions is inconsistence, which suggests that the contact surface between $MoS_2$ and $CoFe_2O_4$ is anisotropic. With the temperature rising from 100 K to 300 K, the two directions in which the magnetoresistance changes with the intensity of the external magnetic field gradually overlap, and the oscillation decreases. In the vertical direction, the two peaks formed by the hysteresis effect also gradually overlap, which is due to the hysteresis effect gradually decreasing with increasing temperature [97].

*3.4 MoTe₂ /EuO heterostructure*

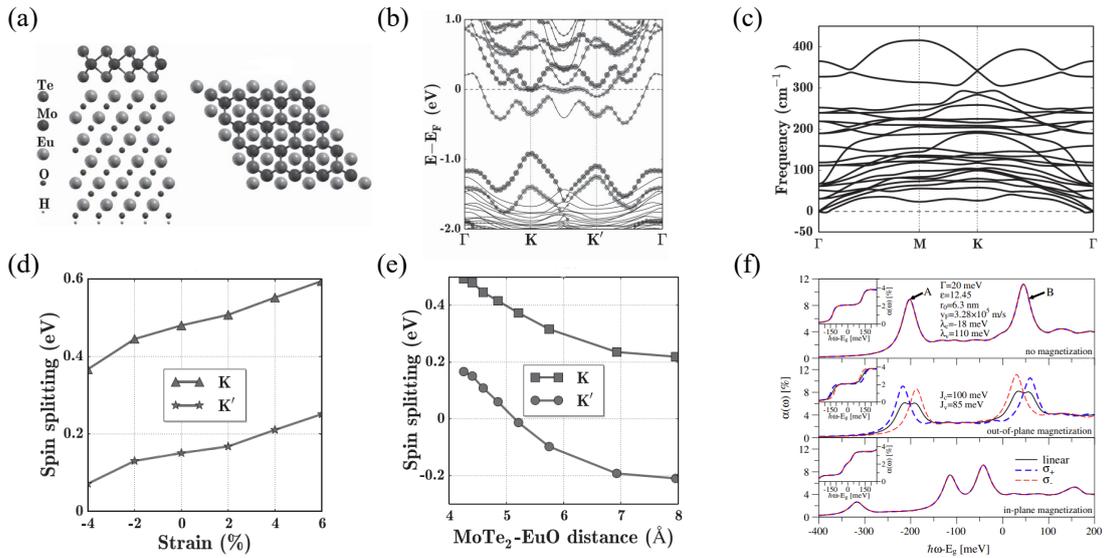

**Fig. 15. Spin-valley polarization of MoS₂ on EuO.** (a) Side view and top view of MoTe₂/EuO structure. (b) Band structure and of the lowest energy configuration of the MoTe$_2$/EuO heterostructure. (c) Phonon spectrum of the lowest energy configuration of monolayer MoTe$_2$ on top of EuO(111). (d) Spin splitting of the two topmost valence bands at the K and K′ valleys as a function of biaxial strain. (e) Spin splitting of the two topmost valence bands at the K and K′ valleys as a function of the MoTe$_2$-EuO distance. (f) The ML absorption spectra simulated by the formula of single-layer MoTe2 under different conditions. Top panel: Spectrum when there is no magnetic proximity effect. Middle panel: Absorption spectrum corresponding to the out-of-plane exchange effect when the magnetic proximity effect is present. Bottom panel: Absorption spectrum corresponding to the in-plane exchange effect when the magnetic proximity effect is present. The magnetic properties of EuO are referenced in this system. Figure adapted with permission from: ref [99], John Wiley and Sons.

The band gap of MoTe$_2$ is 1.0eV (bulk)/1.1eV (single layer) [100]. For a single layer of MoTe$_2$ in MoTe$_2$/EuO, the lattice parameter is a=3.65 Å [98], and the surface mismatch between MoTe$_2$ and EuO is 2.5%. Researchers fixed the lattice constant to the value of the EuO substrate and found a moderate tensile strain is applied to monolayer MoTe$_2$ [99]. DFT calculation points out that the MoTe$_2$/EuO heterostructure (Fig. 15a) has a spin-valley polarization effect and theoretically predicted the nature and source of its self-selected polarization [99]. The band structure obtained for the lowest energy configuration of the MoTe$_2$/EuO heterostructure is shown in Fig. 15b. The finite displacement method is used for the phonon spectrum of the lowest energy configuration to confirm the stability (Fig. 15c) [101-102]. The spin splitting at the two valleys as a function of the MoTe$_2$-EuO biaxial strain (Fig. 15d) and distance (Fig. 15e) reflect the short-range characteristics of the proximity effect. Figure 15f shows the absorption of polarized light of simulated single-layer MoTe$_2$ under different conditions. The figure shows there is no spin polarization in the absence of MPE. When the out-of-plane exchange cleavage is considered, a significant difference in the absorption rate of left and right polarized light could be seen with the introduction of EuO magnetism. It suggests that the MoTe$_2$ film could be magnetized by EuO in MoTe$_2$/EuO via the MPE, thereby triggering the spin valley polarization effect of MoTe$_2$. When the in-plane exchange splitting is considered, no obvious difference in the absorption rate of the left and right polarized light could be seen with the introduction of EuO magnetism, but the A and B peaks are closer than when there is no MPE, and a secondary absorption peak appears from the left, which could be related to the mutual conversion of bright and dark excitons [103].

*3.5 MoSe$_2$ /LSMO heterostructure*

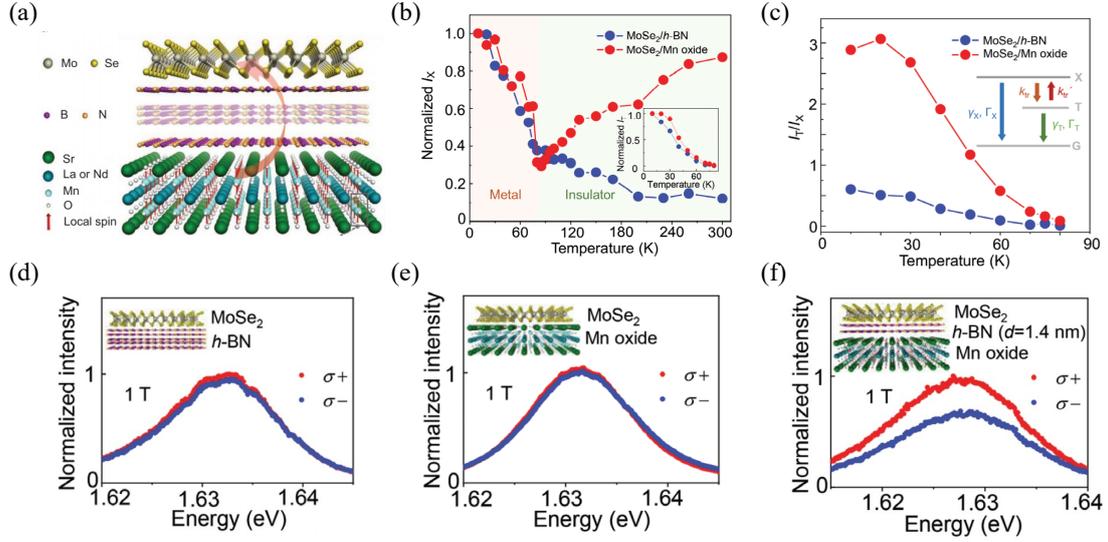

**Fig. 16. Spin polarization of MoSe$_2$ on LSMO** (a) Schematic of the MoSe$_2$/hBN/LSMO. (b) Exciton PL intensity ($I_X$) and trion PL intensity ($I_T$) in MoSe$_2$/hBN and MoSe$_2$/LSMO from 10 to 300 K (c) PL intensity ratio of the trion and exciton ($I_T/I_X$) as a function of temperature from 10 to 80 K. Inset shows the exciton and trion dynamics within a three-level model consisting of the exciton (X), trion (T), and ground state (G). (d-f) Circular polarization-resolved PL spectra for the trion of the reference (1L-MoSe$_2$/h-BN) (d), vdW heterostructure (1L-MoSe$_2$/Mn oxide) (e), and vdW heterostructure (1L-MoSe$_2$/h-BN/Mn oxide, d=1.4 nm) (f) at 10 K under 1 T, right polarized light (σ+) and left polarized light (σ-) were represented by red curve and blue curve respectively. Figure adapted with permission from: ref [104], John Wiley and Sons.

LSMO provides a controllable way to induce MPE for MoSe$_2$. The band gap of MoSe$_2$ is 1.1 eV (bulk)/1.5 eV (single layer) [105]. MoSe$_2$ has a small size effect, quantum size effect and macroscopic quantum effect [106]. Compared with graphene, MoSe$_2$ has a slightly wider band gap, showing better performance in field-effect transistors and low-power electronics. Interlayer hBN's thickness influence on the MPE strength and semiconductor charge transfer degree were studied [104]. The atomic structure diagram of the heterojunction is represented in Fig. 16a. The relationship between the exciton PL intensity $I_X$ and the trion PL intensity $I_T$ in MoSe$_2$/hBN and MoSe$_2$/LSMO with temperature changes (10 K-300 K) has been experimentally studied (Fig. 16 b). The trion PL intensity $I_T$ of MoSe$_2$/hBN and MoSe$_2$/LSMO both decrease with temperature increase. However, the exciton PL intensity $I_X$ of MoSe$_2$/LSMO takes 80 K as the boundary, showing a trend of first decreasing and then increasing. When the temperature is over 80 K, the exciton PL intensity signal changes drastically at the Mn phase transition temperature due to the metal-insulator transition energy gap opening [107]. The exciton kinetic properties of the single-layer MoSe$_2$ can be greatly affected by Mn oxide when the temperature is larger than the phase transition temperature. The Mn oxide controls the exciton state of the single-layer MoSe$_2$ when the temperature is higher than 80 K. Figure 16c shows the integrated PL intensity ratio of the trion and exciton components ($I_T/I_X$) as a function of temperature below $T_C$ (80 K) of Mn oxide.

Notably, $I_T/I_X$ in the heterostructure is significantly larger than that in the reference, where the Mn oxide is in the metallic state. The result suggests that the doped carrier density of MoSe$_2$ on Mn oxide is higher than that of the reference MoSe$_2$ on SiO$_2$/Si. The article further studied the influence of the thickness of hBN on the magnetic neighbors of MoSe$_2$/LSMO. As shown in Fig. 16 d-f, the circular polarization-resolved fluorescence spectrum of vdW heterostructure 1L-MoSe$_2$/h-BN, vdW heterostructure 1L-MoSe$_2$/Mn oxide and vdW heterostructure 1L-MoSe$_2$/h-BN/Mn oxide under 1T external magnetic field and 10 K conditions has been studied experimentally. In the PL spectrum, there is no inconsistency in the left and right circular polarization under the conditions of 1L-MoSe$_2$/hBN and 1L-MoSe$_2$/LSMO, but the left and right circular polarization intensities are different in 1L-MoSe$_2$/hBN/LSMO.

This abnormal phenomenon indicates that MPE does not exist when MoSe$_2$ is directly in contact with LSMO. However, when trilayer hBN is applied between LSMO and MoSe$_2$, MPE will emerge. In consequence, hBN will not shield the MPE between LSMO and MoSe$_2$. On the contrary, trilayer hBN will introduce MPE in this system. The phenomenon is because Mn ions in LSMO have both ferromagnetic and metallic properties. The specific mechanism is shown in Fig. 16f. There is a double exchange interaction between Mn$^{3+}$ and Mn$^{4+}$ in LSMO is the reason for the spontaneous magnetization of LSMO. When MoSe$_2$ is introduced and contact with LSMO directly, an appearance of a metal-induced interstitial state in the contact gap between Mn ions and MoSe$_2$ happens [108, 109]. A new conduction band between Mn ions and MoSe$_2$ will form, and the double exchange effect will vanish[100]. Therefore, LSMO will be non-magnetic because of the direct contact between MoSe$_2$ and LSMO. When 1.4 nm hBN is introduced to separate MoSe$_2$ and LSMO, no interstitial state between LSMO and MoSe$_2$ could be observed. This anomaly allows us to introduce a thin layer of hBN between MoSe$_2$ and LSMO to control the occurrence and shielding of the MPE.

## 4. Conclusion and outlook

In this review, we summarize the state-of-the-art progress of the MPE between 2D materials and various magnetic oxides. Experimentally, the proximity-induced spin splitting in graphene can be revealed by anomalous Hall effect and Zeeman spin Hall effect, while the proximity-induced SOC can be revealed by inverse Rashba-Edelstein effect. XMCD and SPMDS are used to directly probe the proximity-induced magnetism in graphene. Spin-charge conversion, spin transport and spin manipulation have been demonstrated in graphene/magnetic oxide heterostructures. Theoretically, the strength of exchange interaction between graphene and magnetic oxides can be calculated, as shown in Table 1. Because of a strong interaction between graphene and magnetic oxides, the exchange coupling energy can be of the order of tens of meV, and even up to hundreds of meV. The large exchange coupling strength is promising for the design of 2D spintronics and opto-spintronics. Different from graphene-based heterostructures, TMDs on magnetic oxides exhibit various properties such as large band offset and spin valley polarization. Moreover, the strong SOC allows TMDs on magnetic oxides to have optical selectivity. Because of the lack of inversion center in TMDs, hybrid heterostructures between TMDs and magnetic oxides have significant potential in the emergent field of valleytronics.

Over the past years, much progress has been made in graphene-based spin phenomena and devices. Looking forward, there are still many unsolved problems and questions in this field. First, for QAHE, the largest magnitude in proximity graphene can only reach a maximum of ~0.5 $e^2$/h but is not quantized. The weak QAHE is possibly due to the weak SOC strength in graphene. So future works should try to enhance SOC in graphene, for example, by decoration with heavy-metal magnetic adatoms or by hydrogenation. Second, there is a gap between theory and experiment for determining the exchange coupling strength. Theoretical calculations show that the exchange field can be up to tens of meV. However, experimental results show a smaller proximity exchange field. Moreover, different groups have reported different values on the exchange strength even in the same heterostructure. This discrepancy could be ascribed to the various interface qualities, leading to the different interaction strengths. Third, the mechanism of MPE is still an open question. For the same heterostructure, different groups on spin-related measurements have proposed different mechanisms for the results. These differences should prompt additional fundamental studies and spintronics device development in future. TMD materials have become a hot topic since it started being investigated after graphene was isolated. Due to the spin-valley coupling in monolayer TMD material, the spin polarization results in a large valley polarization. Valley and spin hall effect will be achieved in TMD under electric field. Therefore TMD-based heterostructures are competitive candidates for future spintronics.

Finally, the recent development of transfer techniques and moiré physics in 2D materials should give rise to new spin-related phenomena and novel devices through enhanced electron-electron interaction and spin-orbit interaction. Many novel quantum phenomena await to be discovered in the 2D materials/magnetic oxide heterostructures.

**Acknowledgments**

This work is supported by the Ministry of Education (MOE) Singapore under the Academic Research Fund Tier 2 (Grant No. MOE-T2EP50120-0015), by the Agency for Science, Technology and Research (A*STAR) under its Advanced Manufacturing and Engineering (AME) Individual Research Grant (IRG) (Grant No. A2083c0054), and by the National Research Foundation (NRF) of Singapore under its NRF-ISF joint program (Grant No. NRF2020-NRF-ISF004-3518). ATSW acknowledges the Christensen Fellowship at Si Catherine's College, Oxford.

57 (1998) R8079(R).
108. A. Kerelsky, A. Nipane, D. Edelberg, D. Wang, X. Zhou, A. Motmaendadgar, H. Gao, S. Xie, K. Kang, J. Park, J. Teherani, A. Pasupathy, Absence of a band gap at the interface of a metal and highly doped monolayer $MoS_2$, Nano Lett 17 (2017) 5962-5968.
109. J. Tersoff, Schottky barrier heights and the continuum of gap states, Physical Review Letters 52.6 (1984) 465.
32